\documentclass[12pt]{iopart}

\usepackage[dvips]{graphicx}

\begin{document}

\title{Color-flavor locked strangelets and their detection
}

\author{Jes Madsen}

\address{Institute of Physics and Astronomy, University of Aarhus,
DK-8000 {\AA}rhus C, Denmark}

\ead{jesm@ifa.au.dk}

\begin{abstract}
Strange quark matter in a color-flavor locked state is significantly
more bound than ``ordinary'' strange quark matter. 
This increases the likelihood
of strangelet metastability or even absolute stability. Properties of
color-flavor locked strangelets are discussed and compared to ordinary
strangelets. Apart from differences in binding energy, the main
difference is related to the charge. A statistical sample of strangelets
may allow experimental distinction of the two.
Preliminary estimates indicate that the flux of
strangelets in galactic cosmic rays could be sufficient to allow for
strangelet discovery and study in the upcoming Alpha Magnetic
Spectrometer AMS-02 
cosmic ray experiment on the International Space Station.
\end{abstract}

\section{Introduction}

It has long been known that phenomenological strong interaction models,
most notably the MIT bag model, allow absolute stability (energy per
baryon below 930~MeV) of three flavor quark matter for certain ranges of
parameters, and metastability for a wider parameter span. While bag
model calculations are clearly only a crude approximation to full (but
untractable) QCD, the confirmation of strange matter (meta)stability would
have important consequences. 
For instance strange matter stability would imply that
``neutron stars'' are actually quark stars (strange stars), and
metastability could also significantly change the physics of neutron
star interiors. 

Even if quark matter is (meta)stable in bulk, finite size effects
(surface and curvature energies) increase the energy per baryon for
small lumps of three flavor quark matter, called strangelets. This makes
it less likely to form such objects in heavy-ion collisions, and such
formation is also hindered by the high entropy/temperature environment,
which destabilizes strangelets further (``making ice-cubes in a furnace'').
Nevertheless, tiny amounts of strangelets might be formed in colliders,
and the detection of one would be the ultimate smoking gun for the 
quark-gluon plasma \cite{madsen99}.

The probability of strangelet (meta)stability is increased
with the recent demonstration that quark matter at high density
may be in a so-called color-flavor locked phase where quarks 
with different color
and flavor quantum numbers form Cooper pairs with pairing energy
$\Delta$ perhaps as high as 100~MeV \cite{alford01b}. 
Such a state is significantly more bound than ordinary 
quark matter, and this increases the likelihood 
that quark matter composed of up, down, and strange
quarks may be metastable or even absolutely stable. 
In other words color-flavor locked quark matter rather 
than nuclear matter could be the ground state of hadronic matter.

In the following I shall briefly summarize the physics of ``ordinary''
strangelets, followed by a summary of recent work on color-flavor locked
strangelets. Finally, I discuss the potential of experimental discovery
of strangelets with the AMS-02 detector on the International Space
Station, which may even (by studying the mass-charge relation) allow a
distinction between ordinary and color-flavor locked strangelets.

\begin{figure}[h!]
\begin{center}
\includegraphics[height=10cm]{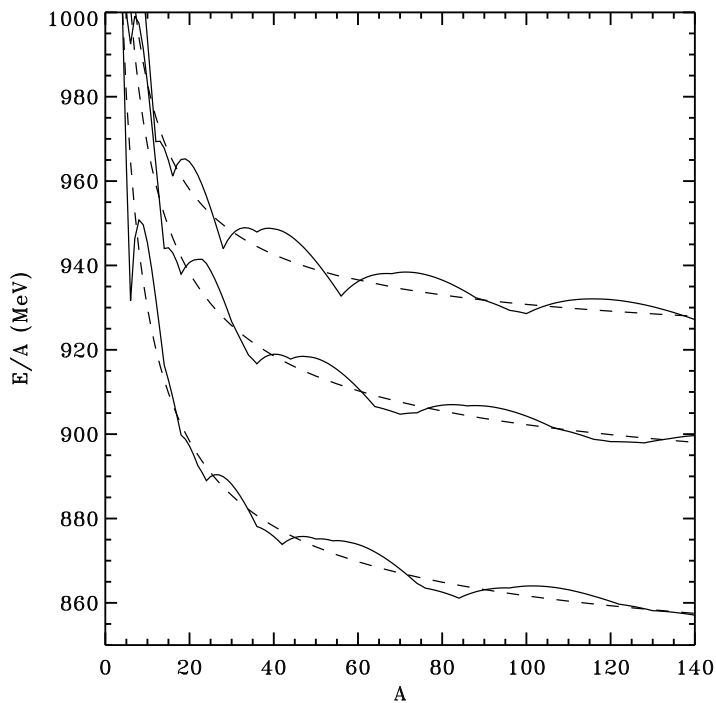}
\caption{Energy per baryon in MeV is plotted as a function of baryon
number for ``ordinary'' strangelets with s-quark mass of
50, 150, and 300~MeV (bottom
to top). The bag constant is $B=(145~{\rm MeV})^4$. Mode filling
within the MIT bag model corresponds to the solid curves, whereas dashed
curves were calculated from the smoothed density of states within the
multiple reflection expansion. It is clearly seen how $E/A$ grows
from a bulk value (c.f.~Fig.~2) at high $A$ to a value significantly
increased by finite size effects at low $A$. It is also seen how the
multiple reflection expansion including volume, surface, and curvature
terms (dashed curves) gives an excellent reproduction of the overall
behavior of $E/A$. The upper set of curves (for $m_s=300$~MeV)
essentially corresponds to two flavor quark matter, since only up- and
down-quarks are energetically favorable for such a high strange quark
mass. The bag constant chosen is close to
the lower bound permitted if nuclei should remain stable against direct
decay into two flavor quark matter (decay into strangelets is forbidden
since it requires a high order weak interaction to create strange
quarks).
}
\label{fig:ordinary}
\end{center}
\end{figure}

\section{Ordinary strangelets}

Strangelet properties are usually studied by filling up, down, and
strange quark energy levels in a spherical bag with MIT bag model
boundary conditions \cite{degrand75}. 
Results of such calculations are shown in Figure 1.
Shell structure reminiscent of nuclear or atomic physics is evident, but
the overall picture is an energy per baryon that decreases from small
to large baryon number $A$, saturating at a bulk value for large $A$.

The general behavior can be understood within a multiple reflection
expansion framework.
Here the energy of a system composed of
quark flavors $i$ is given by
\begin{equation}
E=\sum_i(\Omega_i+N_i\mu_i)+BV,
\label{eq:ordinary}
\end{equation}
where $\Omega_i$, $N_i$ and $\mu_i$
denote thermodynamic potentials,
total number of quarks, and chemical potentials, respectively. $B$ is
the bag constant, $V$ is the bag volume.
The thermodynamical quantities can be derived from a density of states
of the form \cite{balian70}
$
{{dN_i}\over{dk}}=6 \left\{ {{k^2V}\over{2\pi^2}}+f_S\left({m_i\over
k}\right)kS+f_C\left({m_i\over k}\right)C+ .... \right\} ,
$
where a sphere has area $S=4\pi R^2$ and curvature $C=8\pi R$.
For the MIT-bag model
$f_S(m/k)=-\left[ 1-(2/\pi)\tan^{-1}(k/m)\right] /8\pi$ \cite{berjaf87}
and $f_C(m/k)=
\left[ 1-3k/(2m)\left(\pi/2-\tan^{-1}(k/m)\right)\right]/12\pi^2$ 
\cite{madsen94}.
The number of quarks of flavor $i$ is
$
N_i=\int_0^{p_{Fi}}({dN_i}/{dk})dk=n_{i,V}V+n_{i,S}S+n_{i,C}C,
$
and the thermodynamic potentials are
$\Omega_i=\int_0^{p_{Fi}}({dN_i}/{dk})(\epsilon_i(k)-\mu_i)dk=
\Omega_{i,V}V+\Omega_{i,S}S+\Omega_{i,C}C,$
where $\epsilon_i(k)=(k^2+m_i^2)^{1/2}$.
The expressions obey $\partial\Omega_i/\partial\mu_i=-N_i$, and
$\partial\Omega_{i,j}/\partial\mu_i=-n_{i,j}$.
The ground state strangelet with a given $A$ is found by minimizing $E$
with respect to $R$ and $N_i$ for fixed $A$.
As seen from the full lines in Figure 1 such an expansion completely
reproduces the average behavior of $E/A$ versus $A$.

\section{Color-flavor locked strangelets}

\begin{figure}[h!tb]
\begin{center}
\includegraphics[height=10cm]{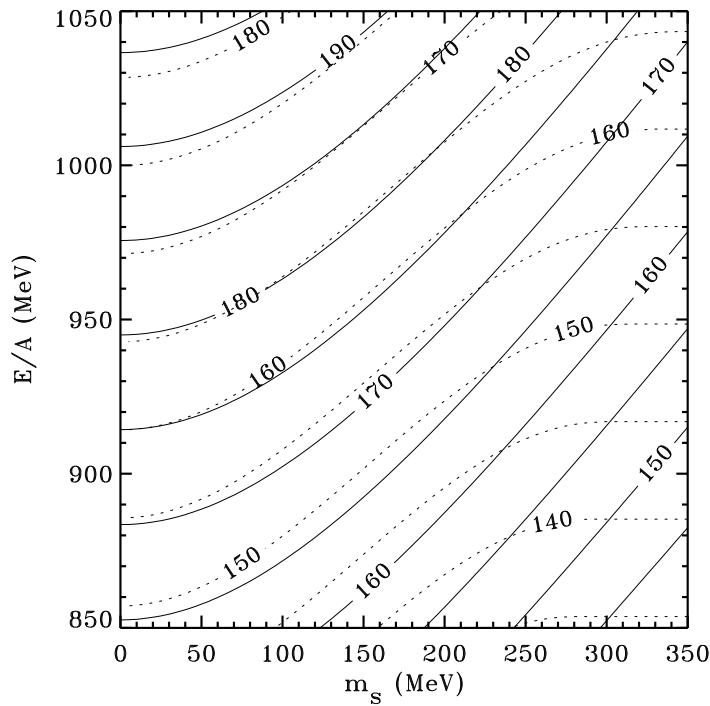}
\caption{Bulk energy per baryon for strange quark matter as a function
of strange quark mass, for several choices of $B^{1/4}$ in MeV. Dotted
curves correspond to ``ordinary'' strange quark matter. Solid curves
include the effect of color-flavor locking with a gap parameter
$\Delta=100$~MeV. It is seen that a typical energy gain for fixed $B$ is
of order 100~MeV per baryon. Notice how $E/A$ for ``ordinary'' strange
quark matter saturates at high $m_s$ for fixed $B$, whereas the
corresponding curve for CFL-strange quark matter continues to rise. This
difference is due to the fact
that increasing $m_s$ makes ordinary quark matter shift from three to
two flavors, whereas the CFL-phase keeps equal numbers of up, down and
strange quarks to maximize the pairing energy even at quite high $m_s$.
}
\label{fig:bulk}
\end{center}
\end{figure}

Strangelets in a color-flavor locked state can be described in a
framework similar to the one summarized above for ordinary strangelets.
This was first attempted in Ref.~\cite{madsen01} (a different approach
to finite size effects in the case of two flavor color superconducting
quark matter was recently presented in \cite{amore01}). In \cite{madsen01} 
the total energy (mass) of a strangelet was written in a manner similar
to Eq.~(\ref{eq:ordinary}) as
\begin{equation}
E=\sum_i (\Omega_i +N_i\mu_i) +(\Omega_{{\rm pair,}V} +B)V ,
\end{equation}
where
$\Omega_{{\rm pair,}V}\approx -3\Delta^2\mu^2/\pi^2$ is the binding
energy from pairing ($\mu$ is the average quark chemical potential), and
the thermodynamic potential of quark flavor $i$ is a sum of volume,
surface, and curvature terms as discussed above.

An important difference relative to
non-CFL calculations is that all quark Fermi momenta in CFL 
strange quark matter are equal. This property leads to
charge neutrality in bulk without any need for electrons
\cite{rajwil01}, and it is due to the
fact that pairing happens between quarks of different color
and flavor, and opposite momenta $\vec p$ and $-\vec p$. It is
therefore
energetically favorable to fill all Fermi seas to the same Fermi
momentum, $p_F$. For bulk quark matter the energy per baryon with and
without color-flavor locking are compared in Fig.~2.
For $\Delta=100$~MeV the gain in energy per baryon is of order 100~MeV
for realistic values of the s-quark mass.

\begin{figure}[h]
\begin{center}
\includegraphics[height=10cm]{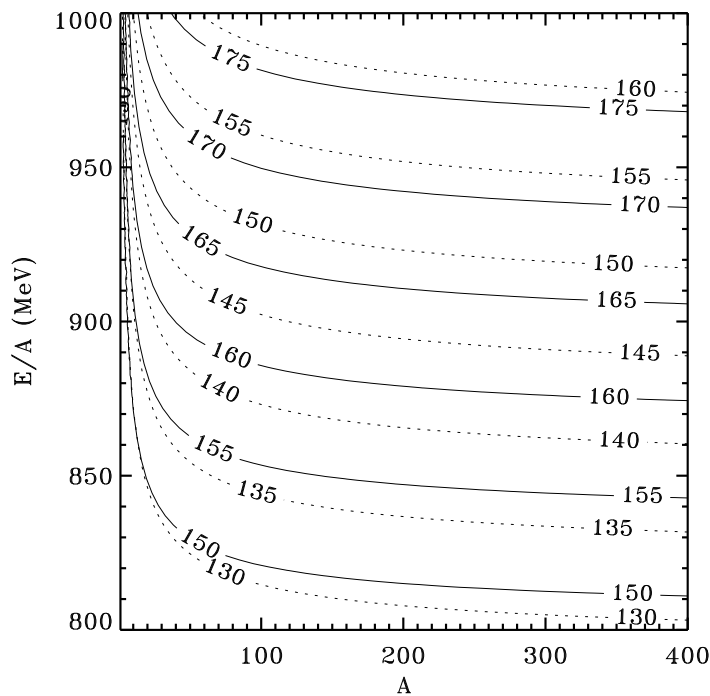}
\caption{Energy per baryon in MeV 
as a function of $A$ for ordinary strangelets
(dotted curves) and CFL strangelets (solid curves) for $B^{1/4}$ in MeV
as indicated, $m_s=150$~MeV, and $\Delta=100$~MeV. All calculations are
performed within the multiple reflection expansion of the MIT bag model.
The dotted curve for $B^{1/4}=145$~MeV corresponds to the middle dashed
curve in Figure 1.}
\label{fig:eovera}
\end{center}
\end{figure}

As illustrated in Figure \ref{fig:eovera}, 
color-flavor locked strangelets have an
energy per baryon, $E/A$, that behaves much like that of ordinary
strangelets as a function of $A$. For high $A$ a bulk value is
approached, but for low $A$ the finite-size contributions from surface
tension and curvature significantly increase $E/A$, making the system
less stable. The main difference from ordinary strangelet calculations
is the overall drop in $E/A$ due to the pairing contribution, which is
of order 100~MeV per baryon for $\Delta= 100$~MeV for fixed values
of $m_s$ and $B$. Since $\Omega_{{\rm pair,}V}\propto \Delta^2$, the
actual energy gain is quite dependent on the choice of $\Delta$.

The charge properties of ordinary strangelets
and CFL strangelets are quite different. Both have a
very small charge per mass unit relative to nuclei, but the exact
relation may provide a way to test
color-flavor locking experimentally if strangelets are found in
accelerator experiments or (perhaps more likely) in cosmic ray
detectors. With $m_{150}\equiv m_s/(150~{\rm MeV})$ 
ordinary strangelets have (roughly)
\cite{farjaf84,berjaf87,heis93}
\begin{equation}
Z\approx 0.1 m_{150}^2 A, ~~~~~~A\ll 10^3,
\end{equation}
\begin{equation}
Z\approx 8 m_{150}^2 A^{1/3}, ~~~~A\gg 10^3 ,
\label{eq:higha}
\end{equation}
whereas CFL strangelets are described by \cite{madsen01}
\begin{equation}
Z\approx 0.3 m_{150} A^{2/3} .
\end{equation}
This relation can easily be understood in terms of the charge neutrality
of bulk CFL strange quark matter \cite{rajwil01} 
with the added effect of the
suppression of s-quarks near the surface, which is responsible for (most
of) the surface tension of strangelets. This leads to a reduced number
of negatively charged s-quarks in the surface layer; thus a total
positive quark charge proportional to the surface area or $A^{2/3}$.

A similar effect becomes important in ordinary
strangelets, meaning that the standard $A^{1/3}$-result, 
Eq.~(\ref{eq:higha}), breaks down at
very high $A$ \cite{madsen00b}. This effect is large enough to
rule out a potential disaster scenario, where negatively charged
strangelets produced in heavy ion colliders could grow by nucleus
absorption and swallow the Earth \cite{jaffe00}. 
While ordinary strange quark matter
can be negatively charged in bulk if the one-gluon exchange $\alpha_S$
is very prominent \cite{farjaf84}, 
the added positive surface charge due to massive s-quark
suppression is sufficient to make the overall quark charge positive for
a large range of $A$, thus preventing any such disaster
\cite{madsen00b}.

Only first steps have been made in the effort to describe
properties of color-flavor locked strangelets, and there is room for
improvement. 
While finite-size effects were included in the free
quark energy calculations, such (unknown) higher order terms were not
taken into account in the pairing energy. This approximation 
is only warranted as long as $\Omega_{\rm pair}$ itself is a perturbation
to $\Omega_{\rm free}$. The discreteness of quark energy levels 
was only
taken into account in an average sense via the smoothed density of
states given as a sum of volume, surface and curvature terms. 
This is an excellent approximation to the average strangelet 
properties (c.f.~Fig.~1) \cite{madsen94}, but it
misses the interesting stabilizing effects near closed shells
\cite{giljaf93,schaff97} that could
make certain baryon number states longer lived than one might expect
from Figure \ref{fig:eovera}. 
Discreteness should also be considered more carefully in the
treatment of pairing for small $A$.
Most important the MIT bag model with $\alpha_S=0$ is only
a crude phenomenological approximation to strong interaction physics;
it is not QCD. 

\section{Strangelet detection at The International Space Station}

While strangelet formation in the
cosmological quark-hadron phase transition seems less likely than
originally believed \cite{witten84,madsen99}, 
a significant flux of cosmic ray
strangelets is expected from another source if strange matter is stable, 
namely collisions of binary 
compact star systems containing strange stars. If strange quark matter is
the ground state of hadronic matter at zero pressure, 
it will be energetically favorable to form strange stars 
rather than neutron stars, and it would be
expected that all the objects normally associated 
with neutron stars (pulsars and low-mass x-ray binaries) 
would actually be strange stars \cite{alcock}.
Several pulsars are observed in binary systems containing another 
compact star. Such binaries move in elliptical orbits,
spiraling closer to each other because the system 
loses energy by gravity wave emission. Ultimately the stars collide.
The expected rate of binary collisions in our Galaxy
is of order $10^{-4}~{\rm year}^{-1}$.

Numerical studies have followed the
late stages of inspiral in systems composed of two 
neutron stars or neutron stars orbiting black holes or white dwarfs.
No detailed calculations have been done for systems containing
strange stars, and since there 
are significant differences in the equation of
state it may be dangerous to rely on existing models. Nevertheless, the
release of a fraction of a solar mass seems to be a generic feature.
Most collisions seem to release between
$10^{-4}$ and $10^{-1}M_\odot$, where $M_\odot$ is the solar mass in
connection with the actual collision and via tidal disruption 
in the late stages of inspiral. 
Lumps of matter released during the
tidal disruption phase are expected to be macroscopic. 
Simple estimates balancing the tidal force
with the surface tension force of strange quark matter
leads to a typical fragment baryon number of order $10^{38}$.

\begin{figure}[h]
\begin{center}
\includegraphics[height=10cm]{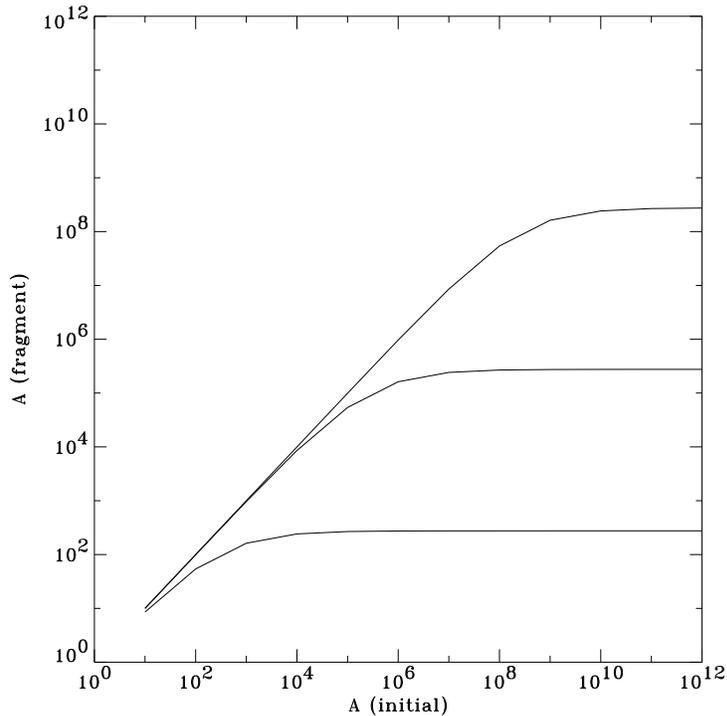}
\caption{The typical $A$-value of strangelet fragments as a function of
the $A$-value of colliding strangelets, for collision energies of
$10^{-6}$, $10^{-4}$, and $10^{-2}$ of the rest mass energy (top to
bottom). Calculations were performed for non-CFL strangelets with
massless quarks and $B^{1/4}=145$~MeV, but similar results are obtained
for other parameter choices. Regardless of initial size, small
lumps in the mass-range detectable from the International Space Station
are abundantly formed for collisions with speed comparable to orbital
speeds in compact binary star systems ($E_{\rm kinetic}\approx 0.01 mc^2$).}
\label{fig:fragment}
\end{center}
\end{figure}

A significant fraction of the tidally released material is 
originally trapped in 
orbits around the binary stars. Typical orbital speeds of the lumps are
$0.1 c$, and collisions among lumps are frequent. Assuming the kinetic
energy in these collisions mainly goes to fragmentation 
of the lumps into smaller
strangelets (i.e.\ that the kinetic energy is used to the extra surface and
curvature energies necessary for forming $N$ lumps of baryon number 
$A/N$ from the original baryon number $A$),
the resulting strangelet distribution peaks at mass numbers from a few
hundred to about $10^3$ as shown in Figure 4. 

This is within the interesting regime for 
the upcoming cosmic ray experiment Alpha Magnetic Spectrometer AMS-02 on
the International Space Station \cite{ams02} (a
prototype AMS-01 was flown on the Space Shuttle mission STS-91 in 1998).
AMS-02 is a roughly 1~m$^2$~sterad detector which will analyze the flux
of cosmic ray nuclei and particles in unprecedented detail for three
years or more following deployment in 2005. It will be sensitive to
strangelets in a wide range of mass, charge and energy \cite{amsstrange}.

Assuming that strangelets share two of the features found 
experimentally for cosmic ray nuclei, namely a
power law energy distribution: $N(E)dE \propto E^{-2.5}$, and an average
confinement time in the galaxy of $10^7$~years, and including a geomagnetic
cutoff rigidity of 6~GeV/c,
the strangelet flux at AMS-02 would be
\begin{equation}
F\approx 5\times 10^5 ({\rm m^2~y~sterad})^{-1}\times R_{-4} \times M_{-2}
\times V_{100}^{-1} \times t_7 ,
\end{equation}
where $R_{-4}$ is the number of strange star 
collisions in our Galaxy per $10^4$
years, $M_{-2}$ is the mass of strangelets ejected per collision in units of 
$10^{-2}M_\odot$, $V_{100}$ is the effective galactic volume in units of
$100~{\rm kpc}^3$ over which strangelets are distributed, and $t_7$ is the
average confinement time in units of $10^7$ years. 
All these factors could be of order
unity if strange matter is absolutely stable, though each with significant
uncertainties.
The flux estimate assumes a charge-mass relation $Z=0.3A^{2/3}$ 
as derived for color-flavor locked strangelets, and is valid for
$A<6\times 10^6$.

Strangelet propagation in the Milky Way Galaxy
is in many ways expected to be similar
to that of ordinary cosmic ray nuclei. Except for a possible background of
slow-moving electrically neutral quark nuggets confined solely by the
gravitational potential of the Galaxy, strangelets are charged 
and are therefore bound to the galactic magnetic field. 
They lose kinetic energy by electrostatic
interactions with the interstellar medium, and they gain energy by Fermi
acceleration in shock waves, for example from supernovae. 
Even if accelerated to relativistic speeds, scatterings on impurities 
in the magnetic field makes the
motion resemble a diffusion process. The solar wind as well as the Earth's
magnetic field become important for understanding the final approach to the
detector. Also, strangelets may undergo spallation 
in collisions with cosmic ray
nuclei, nuclei in the interstellar medium, or other strangelets. 
Much depends on the charge-to-mass relation, 
but the details of propagation are not even well
understood for ordinary nuclei, 
so clearly some uncertainty in the expectations
for the strangelet flux at AMS is inevitable. 

A discovery of strangelets would be a very significant
achievement; an ultimate smoking gun for the quark-gluon
plasma at non-zero chemical potential with profound implications for the
astrophysics of compact stars. 
Experimental information on the charge-to-mass relation may
even allow a test of color-flavor locking in quark matter.
Many uncertainties are clearly
involved in the calculation of the strangelet flux at AMS-02. 
A systematic study
of these issues has been initiated and should significantly improve our
understanding of the strangelet production and propagation. 
But ultimately we
must rely on experiment. It is reassuring, that the simple flux estimates
above lead us to expect a very significant strangelet flux in the AMS-02
experiment.

\ack
This work was supported by The Danish Natural Science Research
Council, and by The Theoretical Astrophysics Center under the Danish
National Research Foundation. I thank Jack Sandweiss and Dick Majka for
discussions and collaboration on strangelet detection with AMS-02.

\end{document}